\documentstyle[12pt]{article}
\begin{document}
\newcommand{\newc}{\newcommand}
\newc{\beq}{\begin{equation}}
\newc{\eeq}{\end{equation}}
\newc{\barr}{\begin{eqnarray}}
\newc{\earr}{\end{eqnarray}}
\newc{\ra}{\rightarrow}
\newc{\lam}{\lambda}
\newc{\eps}{\,\epsilon\,}
\newc{\kap}{\kappa}
\newc{\gev}{\,GeV}
\newc{\half}{\frac{1}{2}}
\newc{\eq}[1]{(\ref{eq:#1})}
\newc{\eqs}[2]{(\ref{eq:#1},\ref{eq:#2})}
\newc{\etal}{{\it et al.}\ }
\newc{\etc}{{\it etc }\ }
\newc{\ie}{{\it i.e.}\ }
\newc{\eg}{{\it e.g.}\ }
\newc{\cG}{{\cal G}}
\newc{\cK}{{\cal K}}
\newc{\Ubar}{{\bar U}}
\newc{\Dbar}{{\bar D}}
\newc{\Ebar}{{\bar E}}
\newc{\nonum}{\nonumber}
\newc{\lab}[1]{\label{eq:#1}}
\newc{\vev}[1]{{<\!{#1}\!>\,}}
\newc{\gsim}{\stackrel{>}{\sim}}
\newc{\lsim}{\stackrel{<}{\sim}}

\title{A New Mechanism to Solve the Small Scale Problem of Local 
Supersymmetry}
\author{A. Chamseddine $^1$, H. Dreiner$^{2,3}$}
\date{{\small $^1$Theoretische Physik, ETH-Z\"urich, CH-8093 
Z\"urich\\
$^2$ Rutherford Lab., Chilton, Didcot, OX11 0QX, UK
\thanks{Permanent Address.}\\
$^3$ Max-Planck-Inst. f\"ur Physik, F\"ohringer Ring 6, D-80805 M\"unchen}}
\maketitle

\vspace{-8.9cm}
\hfill\parbox{8cm}{\raggedleft July, $8^{th}$, 1996\\ RAL-96-048 \\ 
ETH-TH/96-20 \\ MPI-PhT/96-56}
\vspace{8.9cm}

\begin{abstract}
\noindent 
We extend the Standard Model gauge group by a a gauged $U(1)_R$ R-Symmetry or 
a  gauged $U(1)'$. The requirement of cancellation of anomalies is very 
constraining but can be achieved by adding three or four hidden-sector fields
which are Standrad Model singlets. The $U(1)_R$ or $U(1)'$ quantum numbers of 
these singlets are usually large producing a non-renormalisable superpotential
with a high power in the singelt fields. We have minimized the supergravity 
scalar potential and  have found solutions where the vacuum expectation 
values of all hidden-sector singlet fields are less than the Planck mass 
$\vev{z_m}=O(M_{Pl}/10)$. This produces the small supersymmetry scale of 
order the weak scale from only the Planck scale. The mu problem is 
simultaneously solved in this manner.
\vspace{1.7cm}
\end{abstract}


\medskip

One of the basic problems of particle physics is to understand how the 
electroweak scale is generated and why it is so small in comparison with the
Planck scale associated with Newton's constant. In the Standard Model the mass
parameter of the Higgs field suffers from quadratic divergences. The physical
mass parameter must then be tuned to be small, of order the weak-scale, order 
by order in perturbation theory. In supersymmetric theories, it is enough to 
tune the Higgs mass to be small at tree level. However, this does not answer 
the question of how such a small
scale arises in the first place. In locally supersymmetric theories, where
gravity plays an important role, the only natural scale is the Planck
mass. The purpose of this letter is to provide a mechanism which naturally
generates a  small scale of the order of $10^2\gev$ out of the Planck mass
$M_{Pl}=2.43\cdot10^{18}\gev$ without any fine tuning.

The basic idea comes from our work on controlling lepton- and baryon-number
violating operators by imposing an anomaly-free gauged $U(1)'$ or $U(1)_R$  
symmetry \cite{aliherbi1,aliherbi2}, thus giving the leptons and quarks an 
additional gauge quantum number. We work in the framework of local 
supersymmerty which allows for gauging the R-symmetry\cite{freedman}. The most
stringent condition on building such models is that of anomaly cancellation. 
The quantum numbers must also be rational thus guaranteeing that a 
superpotential can be used
to break supersymmetry. It turns out that solutions are difficult to come by.
They typically have a polynomial superpotential with the lowest term having a 
high power in the scalar fields. We then require that minimizing the 
supergravity scalar potential for an appropriate choice of the K\"ahler 
function produces minima for the hidden-sector scalar fields $z_m$ such that
\beq
\vev{\kap z_m}< 1,\quad \forall z_m, \lab{vev}
\eeq
where $\kap=M_{Pl}^{-1}$. The actual values are model dependent. We shall
see below that values of order $\vev{\kap z_m}={\cal O}(0.1)$ will be 
sufficient for our argument. We thus avoid all fine-tuning. For 
sufficiently large powers in $z_m$ we can look for solutions where
\beq
\vev{\kap^2 g(z_m)}=O(m_s).
\eeq
Here $m_s$ is the supersymmetry scale of order $10^2\gev$. One advantage of
this mechanism is that although the superpotential has an infinite number
of terms, only the leading term(s) is (are) relevant.

We extend the minimal supersymmetric Standard Model (MSSM) by a $U(1)_R$ 
gauged R-symmetry or a $U(1)'$ gauge symmetry. The only chiral 
supermultiplets are the
three families of quarks and leptons, $Q_i,\,{\bar D}_i,\, {\bar U}_i,\,L_i,\,
{\bar E}_i$, two pairs of Higgs $SU(2)$-doublets, $H_1,\,H_2,$  and Standard 
Model singlets $z_m$. They have the following gauge quantum numbers 
\barr
L_i&=(1,2,-\half,l_i),\quad\Ebar_i&=(1,1,1,e_i),\quad Q_i=(3,2,\frac{1}{6}
,q_i), \nonum\\
\Ubar_i&=({\bar 3},1,-\frac{2}{3},u_i),\quad\Dbar_i&=({\bar 3},1,\frac{1}{3}
,d_i), \quad H_1=(1,2,-\half,h_1),\nonum \\
H_2&=(1,{\bar 2},\half,h_2),\quad z_m&=(1,1,0,z_m), \lab{quantum}
\earr
with respect to $SU(3)_C\times SU(2)_L\times U(1)_Y\times U(1)_X$. $U(1)_X$ is
either $U(1)_R$ or $U(1)'$. Above, for $U(1)_R$ the R-quantum numbers are 
taken for the fermions, $r^f_i$, which enter into 
the anomaly equations and which we use in the explicit model discussed below.
When minimizing the potential, we shall use the 
bosonic $R$-charges $r^b_i=r^f_i+1$, which are also the charges of the 
superfields. The most general superpotential comprising of a hidden
sector, $z_m$, that breaks supersymmetry and an observable sector, $S_i$, is
\beq
g(z_m,S_i)=g_0(S_i)+g_1(z_m)+g_2(z_m,S_i),
\eeq
where
\beq
g_0(S_i)=h_E^{ij}L_iH_1\Ebar_j + h_D^{ij}Q_iH_1\Dbar_j + h_U^{ij}Q_iH_2
\Ubar_j.
\eeq
$g_2(z_m,S_i)$ contains non-renormalisable interactions mixing the 
hidden-sector with the observable sector, where the only scale that is 
allowed 
to appear is the Planck mass. Imposing the condition that the gauged $U(1)_R$
or $U(1)'$ is anomaly-free determines the possible charges of the fields
$z_m$ and thus the form of the superpotential to be
\beq
g_1(z_m)=\kap^{-3}\sum_{t=1}^\infty a_t (\kap z_1)^{n_{t1}}
 (\kap z_2)^{n_{t2}}\ldots  (\kap z_p)^{n_{tp}}, \lab{superpot}
\eeq
for $p$-hidden sector fields. The sub-powers $n_{ti}$ in the superpotential 
satisfy the gauge invariance constraint
\beq
\sum_{i=1}^p r_i^b n_{ti} = \left\{ \begin{array}{ccc}
2,& {\rm for}& U(1)_R, \\
0,& {\rm for}& U(1)', \end{array}\right. \quad \forall t,
\eeq
where for the $U(1)_R$ case the $r_i^b$ refer to the bosonic charges. We 
shall 
determine solutions where the minimum of the scalar potential satisfies 
Eq.\eq{vev} $\vev{\kap z_m} <1, \quad \forall m$. We shall mainly be 
interested in superpotentials where the total powers of each term
\beq
N_t=\sum_{i=1}^p n_{ti},
\eeq
satisfy 
\beq
{\vev{\kap z_m}}^{N_2}\ll {\vev{\kap z_m}}^{N_1}. \lab{ratio}
\eeq 
Thus the superpotential will be dominated by the first term.

The scalar potential for locally supersymmetric theories is given by
\cite{scalarpot}
\beq
V=\frac{1}{\kap^4}e^{\cG}(\cG^{-1a}_b\cG,_a\cG,^b-3)+\frac{1}{2\kap^4} 
|{\tilde g}_\alpha \cG,_{\alpha}(T^\alpha z)^\alpha|^2,
\eeq
where $\cG$ can be split into a K\"ahler function $\cK$ and a superpotential 
$g$
\beq
\cG=\cK (z^a,z_a)+\ln \frac{\kap^6}{4}|g(z^a)|^2.
\eeq
The D-term of the potential has a Planck size cosmological constant for the 
gauged $U(1)_R$ case \cite{freedman}. Therefore the anlyses for the $U(1)_R$ 
and the $U(1)'$ case are not identical and we consider them separately.

{\bf 1.}{\it Gauged $U(1)_R$ Case}\newline
The choice of the K\"ahler function is only restricted by the physical 
requirement that at low energies ($\kap\ra0$) the kinetic energy of 
the scalar fields
\beq
\frac{1}{\kap^2} \cK,_a^b D_\mu z^a D^\mu z_b,
\eeq
becomes canonical. The simplest possibility is to take a universal $\cK=
\cK(u)$ 
where $u=z_az^a$ (summation over $a$), which includes the case of minimal
kinetic energy, $\cK(u)=\kap^2 u$. The minimum of the  scalar potential
should satisfy the properties
\barr
V,_a&=&0, \\
V&=&0, \\
(D-term) &=& 0.
\earr
One can show, however, that it is {\it not} possible to satisfy these 
properties simultaneously when the K\"ahler function has the universal form 
$\cK=\cK(u)$ and the hidden-sector potential is dominated by a single leading 
term. We thus consider the more general form for the K\"ahler function
\beq
\cK=\sum_a \cK^{(a)}(u^{(a)}),\quad u^{(a)}=z_az^a.\label{kahl1}
\eeq
We now determine the minimum of the scalar potential for the condition
\eq{ratio}. We thus consider only the leading term of the hidden-sector
superpotential. Perturbative corrections to the minimum from higher terms will
be small. Therefore the local minimum which we determine 
is stable. However, the question remains whether the complete potential has a 
different {\it global} minimum. At present, this potential is intractable and 
far beyond the scope of this letter. Given these assumptions, the potential 
for the hidden-sector simplifies to
\barr
V&=&\frac{\kap^2}{4}|g|^2e^{\Sigma\, \cK^{(a)}}\left[\sum_a F(u^{(a)})-3
\right] +\frac{{\tilde g}_R^2}{18\kap^4}|(\cK'ru)^{(a)}+2|^2,
\earr
where $(\cK'ru)^{(a)}=\cK'^{(a)} r^{(a)} u^{(a)}$ and the function $F$ for
a given $u^{(a)}$ is given by 
\beq
F(u) =\frac{u}{\cK'+u\cK''} (\cK'+\frac{n}{u})^2.
\eeq
Here $n$ is the power of $u$ in the leading term of the hidden-sector
superpotential. At the minimum, we must then have 
\barr
F'(u^{(a)})&=&0,\qquad \forall u^{(a)}, \\
\sum_a F(u^{(a)}) &=& 3, \\
\sum_a (\cK'ru)^{(a)}&=&-2.
\earr
It is difficult to find functions $\cK(u^{(a)})$ satisfying all these 
constraints
and maintaining the positivity of the scalar kinetic energy. One appropriate 
K\"ahler function is a deformation  of the no-scale function first used in
\cite{freedman2}
\beq
\cK^{(a)}(u^{(a)})=\frac{n_a}{\beta_a} \ln(1+\alpha\sigma^2 n_a u^{(a)}),
\lab{kahl}
\eeq
where $\sigma$ and $\beta_a$ are parameters, $\alpha=\pm1$. We have inserted 
$\sigma^2$ since it is $\sigma$ which sets the scale of $\vev{z_m}$ ($u^{(a)}
=z_az^a$). The conditions at the minimum of the scalar potential are satisfied
provided that
\barr
\sum_{a=1}^p n_a(1+\beta_a) &=& \frac{3}{4},\lab{eq1}\\
\sum_{a=1}^p \frac{r^b_an_a}{1+2\beta_a}&=& -2,\\
u^{(a)}&=& \frac{\alpha}{\sigma^2 n_a}\frac{\beta_a}{1+\beta_a}. \lab{eq3}
\earr
The kinetic energy of the scalar fields is positive if ${\alpha}/{\beta_a}>0$.
Since $u^{(a)}\geq0$ we have two sets of solutions: if $\alpha=+1$ then 
$\beta_a>0$, if $\alpha=-1$ then $-1<\beta_a<0$. It is straight forward to 
show that the only solution to these equations with these conditions is for 
all $\beta_a\eps]-1,0[$. The mass scale which is related to the value of the 
superpotential at the minimum is \cite{susyscale}
\beq
m_s^2=(\kap^2g')^2= \frac{1}{\kap^2}a_1^2\cdot (\kappa^2u_1)^{n_1}\cdots 
(\kappa^2u_p)^{n_p} . \lab{scale}
\eeq

For illustration we give the following anomaly-free model based on our
previous work \cite{aliherbi2} where the full anomaly-equations were given for
the case of family-dependent charges. Consider a left-right symmetric model
where the $U(1)_R$ charges of Eq.\eq{quantum} satisfy: $e_i=l_i$, and $d_i=u_i
=q_i$. Furthermore, assume dominant third generation Yukawa couplings, so that
only $h_E^{33},h_D^{33},h_U^{33}\not=0$ in the superpotential at tree-level. A
possible solution for the fermionic charges $r^f_i$ is 
\beq
(\{l_1,l_2,l_3\};\{q_1,q_2,q_3\};\{h_1,h_2\})=
(\{-\frac{11}{2},-8,0\};\{-\frac{47}{6},\frac{28}{3},0\};\{-1,-1\})
\eeq
The remaining two anomaly equations containing the singlets are
\cite{aliherbi2}
\beq
\sum_m z_m^3 =\frac{23518}{8}, \quad
\sum_m z_m=\frac{45}{2}.
\eeq
For three singlets these equations have many solutions. The power of the 
leading hidden-sector superpotential term can vary from 3 to well over 100 
with
a continuous scatter. As an example, we consider the solution $(z_1,z_2,z_3)
=(-\frac{75}{2}, \frac{57}{2},\frac{63}{2})$. The lowest power of the 
superpotential is then $N_1=20$, with $(n_1,n_2,n_3)=(9,9,2)$. The next term 
has $N_2=26$ which is sufficiently suppressed. We have solved 
Eqs\eq{eq1}-\eq{eq3} in terms of $\beta_3$ and we find a continous set of 
solutions with $\beta_3\eps ]-1,-0.83]$. A specific solution is
\beq
\beta_1=-0.96, \quad \beta_2=-0.97, \quad \beta_3=-0.91.
\eeq
We then obtain for the supersymmetry scale \eq{scale}
\beq
m_s={{a_1}}{\sigma^{-20}}\cdot(2.4\cdot10^{24} )\gev 
\eeq
which is of order the weak scale for $\vev{z}\sim\frac{1}{\sigma}=0.075
-0.085$, $a_1={\cal O}(1)$. The mu-term $\mu H_1H_2$ is prohibited in the 
superpotential at tree-level. The effective mu-term generated from $b_1\kap
\prod (\kap z_i)^{n_i}H_1H_2$ has the same suppression as the leading term 
in the superpotential and therefore $\mu_{eff}= {\cal O}(m_s)$, for $b_1=
{\cal O}(1)$.

\medskip

{\bf 2.}{\it Gauged $U(1)'$ Case}\newline
The advantage of the $U(1)'$ case is that one can make use of a universal
K\"ahler function
\beq
\cK=\cK(u),\quad u=z_az^a.
\eeq
As before, we look for solutions where the superpotential is dominated by the 
first term. The scalar potential then reduces to
\beq
V=\frac{\kap^2}{4} e^\cK |g|^2\frac{1}{K'} \left[ \sum \frac{n_a^2}{u_a}
+F(u) \right] + \frac{1}{8\kap^4}
\cK'^2{\tilde g}'^2 |q_a u_a|^2,
\eeq
where $u_a=z_az^a$ (no summation), $u=\sum u_a$ and now
\beq
F(u)= \frac{\cK'}{\cK'+u\cK''}(\cK'^2u -N^2\frac{\cK''}{\cK'}+2N\cK) -3\cK'.
\lab{feq}
\eeq
Here $N=\sum n_a$.
Imposing the conditions that the $D$-term vanishes at the minimum we find that
the total potential vanishes at the minimum.  The remaining conditions are
\barr
u_a&=& \frac{n_a}{N}u,\\
F&=&-\frac{N^2}{u},\\
F'&=&\frac{N^2}{u^2}.\lab{fpeq}
\earr
We choose the same form for the K\"ahler function $\cK$ as in Eq.\eq{kahl}
\beq
\cK=-\frac{\lambda}{c} \ln(1-cu). \lab{newkahl}
\eeq
The equations \eq{feq}-\eq{newkahl} can then be reduced to one linear 
equation in $u$
\barr
u\left(\frac{1}{N^2c^2} \cdot A\left\{ \frac{1}{N^2c^2}(\lam-Nc)^2\cdot A
+B\right\}  \right. &&\left.\hspace{-1cm}+ N^2c-\frac{1}{c^2} (\lam-Nc)^2 
\right)
\nonum \\&=& N^2+\frac{1}{c^2} \left( \frac{1}{N^2c^2}(\lam-Nc)^2\cdot A +B 
\right)
\lab{linu}
\earr
where
\beq
A= (2N^2c+3\lam c-\lam^2), \quad B=(2N\lam-N^2c-3\lam).
\eeq
The low-energy mass scale is generated by the expectation value of the 
hidden-sector superpotential
\beq
m_s^2=(\kap^2 g_1)^2= a_1^2 (n_1)^{n_1} (n_2)^{n_2} \ldots (n_p)^{n_p} 
(\frac{1}{N})^N u^N (\frac{1}{\kap^2}). \lab{susyscale2}
\eeq
For illustration we give the following model. We consider family-independent 
$U(1)'$ charges, \ie $l_1=l_2=l_3=l$, {\it etc}. The anomaly-equations are 
given
in our notation in \cite{aliherbi1}. We employ the Green Schwarz mechanism 
\cite{greenschwarz} to cancel the anomalies and thus include additional 
constants $c_1,c_2,c_3$ in the anomaly-equations as for example in 
\cite{aliherbi2}. As a possible solution, we find
\barr
(l,e,q,u,d,h_1,h_2) &=&(1,-1,-10,-14,10,0,24), \\
(c_1,c_2,c_3)&=&(-\frac{81}{2},-\frac{63}{2},-36). \nonum
\earr
The remaining anomaly equations are
\beq
\sum_m z_m^3 = 6045,\quad \sum_m z_m = 165.
\eeq
For four singlets we have found only two solutions. The solution with the 
higher leading superpotential power is $(z_1,z_2,z_3,z_4) = (-193,131, 168, 
59),$ with $N_1=14$ and $(n_1,n_2,n_3,n_4) \newline
= (5,3,2,4)$. $N_2=17$. Inserting 
$N_1=14$ into Eq\eq{linu} we obtain a set of solutions for $u$. 
As an example solution, we present
\beq
\lam=7.8,\quad c=-2.15,\quad \Rightarrow \quad u=0.012.
\eeq
The supersymmetry scale \eq{susyscale2} is of order the weak scale for $a_1=
10$. The mu-term is again prohibited at tree-level. The lowest power term $b_1
\kap \prod(\kap z_i)^{m_i} H_1H_2$ generating an effective term $\mu H_1
H_2$ has a power $\sum m_i=12$ and $(m_i)=(3,1,0,8)$. The mass parameter
$\mu$ is then of order the weak scale for $b_1=1/20$.

\medskip

In conclusion, within the framework of local supersymmetry, we have 
demonstrated a mechanism which dynamically breaks supersymmetry and generates 
a supersymmetry scale of order the weak scale from the Planck scale alone,
without any fine-tuning. We have minimally extended the supersymmetric 
standard model by an anomaly-free  $U(1)$ gauge symmetry which can be an 
$R$-symmetry as well as three to four Standard Model singlets. The mu-problem
is simultaneously solved by the same mechanism.


\begin{thebibliography}{99}
\bibitem{aliherbi1}{A.H. Chamseddine, H. Dreiner, Nucl. Phys. B 447 (1995) 
195; hep-ph/9503454.}
\bibitem{aliherbi2}{A.H. Chamseddine, H. Dreiner, Nucl. Phys. B  458 (1996) 
65; hep-ph/9504337.}
\bibitem{freedman}{D.Z. Freedman, Phys. Rev. D 15 (1977) 1173; S. Ferrara, L.
Girardello, T. Kugo and A. van Proeyen, Nucl. Phys. B 223 (1983) 191.}
\bibitem{scalarpot}{E. Cremmer, S. Ferrara, L. Giradello and A. van Proeyen,
Nucl. Phys. B. 212, (1983) 413.} 
\bibitem{freedman2}{D. Castano, D.Z. Freedman, and C. Manuel, Nucl. Phys. B
461 (1996) 50; hep-ph/9507397.}
\bibitem{susyscale}{A. H. Chamseddine, P. Nath and R. Arnowitt, 
Phys. Rev. Lett, 49 (1982) 970; R. Barbieri, S. Ferrara and C. Savoy, Phys. 
Lett B 119 (1982) 343.}
\bibitem{greenschwarz}{M. B. Green and J. H. Schwarz, Phys. Lett. B 149 (1984)
117. }
\end{thebibliography}
\end{document}